\documentstyle[preprint,revtex]{aps}
\begin {document}
\draft
\preprint{UCI TR 92-24 /Uppsala U. PT14 1992}
\begin{title}
Screening of Very Intense Magnetic Fields by\\ Chiral
Symmetry Breaking
\end{title}
\author{Myron Bander\footnotemark\ }
\begin{instit}
Department of Physics, University of California, Irvine, California
92717, USA
\end{instit}
\author{H. R. Rubinstein\footnotemark\ }
\addtocounter{footnote}{-1}\footnotetext{e-mail:
mbander@ucivmsa.bitnet;
mbander@funth.ps.uci.edu}\addtocounter{footnote}{1}%
\footnotetext{e-mail:
rub@vand.physto.se}
\begin{instit}
Department of Radiation Sciences, University of Uppsala, Uppsala,
Sweden
\end{instit}
\receipt{May\ \ \ 1992}
\begin{abstract}
In very intense magnetic fields, $B > 1.5\times 10^{14}$ T, the
breaking of the strong interaction $SU(2)\times SU(2)$ symmetry
arranges itself so that instead of the neutral $\sigma$ field acquiring
a vacuum expectation value it is the charged $\pi$ field that does and
the magnetic field is screened. Details are presented for a magnetic
field generated by a current in a wire; we show that the magnetic
field is screened out to a distance $\rho_o\sim I/f_\pi m_\pi$ from the
wire.
\end{abstract}
\newpage
\narrowtext

It has been recently recognized that fields with complicated
interactions of non-electromagnetic origin can induce various
instabilities in the presence of very intense magnetic fields. By very
intense we mean $10^{14}$ T to $10^{20}$ T. Fields with anomalous
magnetic moments \cite{Ambjorn} or fields coupled by transition
moments \cite{BanRub1} induce vacuum instabilities.  For fields greater
than $5\times 10^{14}$ T the proton becomes heavier than the neutron
and decays into the latter by positron emission \cite{BanRub2}.  In
this work we show that the usual breaking of the strong interactions
chiral symmetry is incompatible with very intense magnetic fields.
Using the standard $SU(2)\times SU(2)$ chiral $\sigma$ model we show
that magnetic fields $B\ge B_c$ with $B_c={\sqrt 2}m_{\pi}f_{\pi}$ are
screened; $f_{\pi}=132$ MeV is the pion decay constant and $m_{\pi}$ is
the mass of the charged pions. This result is opposite to what
occurs in a superconductor; in that case it is weak fields that are
screened and large ones penetrate and destroy the superconducting state.

As the magnetic fields are going to be screened we must be very
careful in how we specify an external field. One way would be to give
$f_{\pi}$ a spatial dependence and take it to vanish outside some
large region of space. In the region that $f_\pi$ vanishes we could
specify the external
field and see how it behaves in that part of space where chiral
symmetry is broken. This is the procedure used in studying the
behavior of fields inside superconductors. In the present situation we
find this division artificial and, instead of specifying the magnetic
fields, we shall specify the external currents. Specifically we will
look at the electromagnetic field coupled to the charged part of
the $\sigma$ model and to the current $I$ in a long straight wire.
{}From this result it will be easy to deduce the behavior in other
current configurations.

The Hamiltonian density for this problem is
\begin{eqnarray}
H=&&{1\over 2}\mbox{\boldmath$\nabla$}\sigma\cdot \mbox{\boldmath
$\nabla$}\sigma + {1\over 2}\mbox{\boldmath $\nabla$}\pi_0\cdot
\mbox{\boldmath $\nabla$}\pi_0+ (\mbox{\boldmath $\nabla$}+e{\bf
A}){\pi}^{\dag}\cdot(\mbox{\boldmath $\nabla$}-e{\bf A}){\pi}\nonumber\\
+&&g(\sigma^2+{\bf\pi}\cdot{\bf\pi}-f^2_{\pi})^2+m_{\pi}^2(f_{\pi}-
\sigma) +
{1\over 2}(\mbox{\boldmath $\nabla\times A$})^2 -
 {\bf j}\cdot {\bf A}\, ;
\end{eqnarray}
$\bf j$ is the external current.
We have used cylindrical coordinates with \mbox{\boldmath$\rho$}
the two dimensional vector normal to the $z$ direction. We will study
this problem in the limit of very large $g$, where the radial degree
of freedom of the chiral field is frozen out and we may write
\begin{eqnarray}
\sigma&&=f_{\pi}\cos\chi\, ,\nonumber\\
\pi_0&&=f_{\pi}\sin\chi\cos\theta\, ,\nonumber\\
\pi_x&&=f_{\pi}\sin\chi\sin\theta\cos\phi\, ,\nonumber\\
\pi_y&&=f_{\pi}\sin\chi\sin\theta\sin\phi\, .\label{angvar}
\end{eqnarray}
In terms of these variables the Hamiltonian density becomes
\begin{eqnarray}
H=&&{{f^2_{\pi}}\over 2}(\mbox{\boldmath$\nabla$}\chi)^2+
{{f^2_{\pi}}\over 2}\sin^2\chi(\mbox{\boldmath$\nabla$}\theta)^2+
{{f^2_{\pi}}\over2}
\sin^2\chi\sin^2\theta(\mbox{\boldmath$\nabla$}\phi- e{\bf A})^2
\nonumber\\
+&&m^2_{\pi}f^2_{\pi}(1-\cos\chi)+
{1\over 2}(\mbox{\boldmath $\nabla\times A$})^2
-{\bf j}\cdot {\bf A}\, .\label{hamdens}
\end{eqnarray}
The angular field $\phi$ can be eliminated by a gauge transformation.
For a current along a long wire we have
\begin{equation}
{\bf j}=I\delta(\mbox{\boldmath $\rho$}){\bf z}\, ;
\end{equation}
The vector potential will point along the $z$ direction, ${\bf
A}=A{\bf z}$ and the
fields will depend on the radial coordinate only. The equations of
motion become
\begin{eqnarray}
-\mbox{\boldmath$\nabla$}^2\chi+
\sin\chi\cos\chi(\mbox{\boldmath$\nabla$}\theta)^2+
e^2\sin\chi\cos\chi\sin^2\theta A^2+m^2_{\pi}\sin\chi=&&0\, ,\nonumber\\
\mbox{\boldmath$\nabla$}(\sin^2\chi\mbox{\boldmath$\nabla$}\theta)
+e^2\sin^2\chi
\sin\theta\cos\theta A^2=&&0\, ,\nonumber\\
-\mbox{\boldmath$\nabla$}^2A+e^2f^2_{\pi}\sin^2\chi\sin^2\theta
-I\delta(\mbox{\boldmath $\rho$})=&&0\, .\label{eqmot}
\end{eqnarray}
In the absence of the chiral field the last of the Eqs.\
(\ref{eqmot}) gives the classical vector potential due to a long wire
\begin{equation} A={I\over 2\pi}\ln{\rho\over a}\, ,
\end{equation}
with $a$ an ultraviolet cutoff. The energy per unit length in the
$z$ direction associated with this configuration is
\begin{equation}
E={I^2\over 4\pi}\ln{R\over a}\, ,\label{classener}
\end{equation}
where $R$ is the transverse extent of space (an infrared cutoff).

Before discussing the solutions of (\ref{eqmot}) it is instructive to
look at the case where there is no explicit chiral symmetry breaking,
$m_{\pi}=0$. The solution that eliminates the infrared divergence in
Eq.\ (\ref{classener}) is $\chi=\theta=\pi /2$ and $A$ satisfying
\begin{equation}
-\mbox{\boldmath$\nabla$}^2A+e^2f^2_{\pi}A
-I\delta(\mbox{\boldmath $\rho$})=0\, .
\end{equation}
For any current the field $A$ is damped for distances $\rho >
1/ef_{\pi}$ and there is no infrared divergence in the energy. (Aside
from the fact that chiral symmetry is broken explicitly, the reason
the above discussion is only of pedagogical value is that the
coupling of the pions to the quantized electromagnetic field does break
the $SU(2)\times SU(2)$ symmetry into $SU(2)\times U(1)$ and the charged
pions get a light mass, $m_{\pi}\sim 35$ MeV \cite{35pi}, even in the
otherwise chiral symmetry limit.)

The term in Eq.\ (\ref{hamdens}) responsible for the pion mass prevents
us from setting $\chi=\pi/2$ everywhere; the energy density would behave
as $\pi f^2_{\pi}m^2_{\pi}R^2$, an infrared divergence worse than that
due to the wire with no chiral field present. We expect that $\chi$
will vary from $\pi/2$ to $0$ as $\rho$ increases and that
asymptotically we will recover classical electrodynamics. Although we
cannot obtain a closed solution to Eqs.\  (\ref{eqmot}), if the
transition between $\chi=\pi/2$ and $\chi=0$ occurs at large $\rho$,
we can find an approximate solution. The approximation consists of
neglecting the $(\mbox {\boldmath$\nabla$}\chi)^2$ term in Eq.\
(\ref{hamdens}); we shall return to this shortly. The solution of
these approximate equations of motion is
\begin{eqnarray}
\chi&&=\left\{\begin{array}{ll}
              {\pi\over 2}\ \   & \mbox{for $\rho < \rho_0$}\\
              0     &   \mbox{for $\rho > \rho_0$}\, ,
             \end{array} \right. \nonumber\\
\theta&&={\pi\over 2}\, , \nonumber\\
A&&=\left\{\begin{array}{ll}
     -{I\over 2\pi}\left[ K_0(ef_{\pi}\rho)-
     {{I_0(ef_{\pi}\rho)K_0(ef_{\pi}\rho_0)}/
     I_0(ef_{\pi}\rho_0)}\right]\ \  & \mbox{for $\rho <\rho_0$}\\
   {I\over 2\pi}\ln{\rho\over\rho_0}  & \mbox{for $\rho >\rho_0$}\, ;
    \end{array} \right.
\end{eqnarray}
$\rho_0$ is a parameter to be
determined by minimizing the energy density of Eq.\  (\ref{hamdens}).
Note that for $\rho > \rho_0$ the vector potential as well as the
field return to values these would have in the absence of any chiral
fields and that for $\rho < \rho_0$ the magnetic field decreases
exponentially as $B\sim \exp (-ef_{\pi}\rho)$. The physical picture is
that, as in a superconductor, near $\rho=0$ a cylindrical current sheet
is set up that opposes the current in the wire and there is a return
current near $\rho=\rho_0$; Amp\`{e}re's law insures that the field at
large distances is as discussed above.
The energy density for the above configuration, neglecting
the spatial variation of $\chi$, is
\begin{equation}
H=-{I^2\over 4\pi}\left [{K_0(ef_{\pi}\rho_0)\over
I_0(ef_{\pi}\rho_0)}+\ln(ef_{\pi}\rho_0)\right ] +
\pi m^2_\pi f^2_\pi \rho^2 +\cdots\, ,\label{apprhamdens}
\end{equation}
where the dots represent infrared and ultraviolet regulated terms
which are, however, independent of $\rho_0$. For $\rho_0>1/ef_\pi$
the term involving the Bessel functions may be neglected and
minimizing the rest with respect to $\rho_0$ yields
\begin{equation}
\rho_0={I\over 2{\sqrt 2}\pi m_\pi f_\pi}\, .\label{rho0}
\end{equation}
This is the main result of this work.

We still have to discuss the validity of the two approximations we
have made. The neglect of the Bessel functions in Eq.\
(\ref{apprhamdens}) is valid for $ef_{\pi}\rho_0 > 1$ which in
turn provides a condition on the current $I$, $eI/m_{\pi} > 2{\sqrt
2}\pi$ or more generally
\begin{equation}
I/m_{\pi} >> 1\, . \label{condition}
\end{equation}
The same condition permits us to
neglect the spatial variation of $\chi$ around $\rho=\rho_0$. Let
$\chi$ vary from $\pi/2$ to $0$ in the region $\rho-d/2$ to
$\rho+d/2$, with $1/d$ of the order of $f_{\pi}$ or $m_{\pi}$. The
contribution of the variation of $\chi$ to the energy density is
$\Delta H=\pi^3f^2_{\pi}\rho_0 d$. Eq.\  (\ref{condition}) insures that
$\Delta H$ is smaller than the other terms in Eq.\
(\ref{apprhamdens}).

Eq.\ (\ref{rho0}) has a very straightforward explanation. It results
from a competition of the magnetic energy density ${1\over 2}B^2$ and
the energy density of the pion mass term $m^2_\pi f^2_\pi
(1-\cos\chi)$.  The magnetic field due to the current $I$ is $B={I/
2\pi\rho}$ and the transition occurs at $B=B_c$, with $B_c={\sqrt
2}m_\pi f_\pi$.  Thus, for any current configuration, the chiral
fields will adjust themselves to screen out fields larger than $B_c$.
Topological excitations may occur in the form of magnetic vortices;
the angular field $\phi$ of Eq.\ (\ref{angvar}) will wind around a
quantized flux tube of radius $1/ef_\pi$ \cite{FetWal}.

Are there situations where magnetic fields of such magnitudes might be
present?  The value of the critical field discussed above is $B_c\sim
1.5\times 10^{14}$ T. Possible astrophysical phenomena where fields of
such magnitudes may occur have been discussed in Refs.\
\cite{Ambjorn,BanRub1,BanRub2}. Superconducting cosmic strings
\cite{Witten} may carry currents $I=10^{20}$ A in a ``wire'' of
thickness $1/M_W$; the magnetic field would be screened out to a
distance of $40$ cm.

M.\ B.\ was supported in part by the National Science Foundation under
Grant No.\ PHY-89-06641. H.\ R. was supported by a SCIENCE EEC
Astroparticles contract.

\end{document}